\documentclass[runningheads]{llncs}
\usepackage{graphicx}
\usepackage{xcolor}

\usepackage{hyperref}

\usepackage{booktabs}

\usepackage{multirow}

\usepackage{subcaption}

\begin{document}
%
%
%
%
%
%







%

\title{
Learning Generic Lung Ultrasound Biomarkers for Decoupling Feature Extraction from Downstream Tasks
}

\titlerunning{Learning Generic Lung Ultrasound Biomarkers}
%
%

\author{
Gautam Rajendrakumar Gare\inst{1}\thanks{corresponding author gautam.r.gare@gmail.com; Confidential copy; Under review at MICCAI 2022.} \and
Tom Fox\inst{2} \and 
Pete Lowery\inst{2} \and 
Kevin Zamora\inst{2} \and 
Hai V. Tran\inst{2} \and 
Laura Hutchins\inst{2} \and 
David Montgomery\inst{2} \and 
Amita Krishnan\inst{2} \and 
Deva Kannan Ramanan\inst{1} \and 
Ricardo Luis Rodriguez\inst{3} \and
Bennett P deBoisblanc\inst{2} \and 
John Michael Galeotti \inst{1}
}

\authorrunning{G. Gare et al.}
\institute{Robotics Institute, Carnegie Mellon University, USA \and
Dept. of Pulmonary and Critical Care Medicine, Louisiana State University, USA \and
Cosmeticsurg.net, LLC, Baltimore, USA\\
}

\maketitle              
%


\begin{abstract}
Contemporary artificial neural networks (ANN) are trained end-to-end, jointly learning both features and classifiers for the task of interest. Though enormously effective, this paradigm imposes significant costs in assembling annotated task-specific datasets and training large-scale networks. We propose to decouple feature learning from downstream lung ultrasound tasks by introducing an auxiliary pre-task of visual biomarker classification. We demonstrate that one can learn an informative, concise, and interpretable feature space from ultrasound videos by training models for predicting biomarker labels. Notably, biomarker feature extractors can be trained from data annotated with weak video-scale supervision. These features can be used by a variety of downstream \textit{Expert} models targeted for diverse clinical tasks (Diagnosis, lung severity, S/F ratio). Crucially, task-specific expert models are comparable in accuracy to end-to-end models directly trained for such target tasks, while being significantly lower cost to train.

\keywords{Lung US Biomarkers \and Transfer Learning \and Diagnostic Tasks}
\end{abstract}
%
%
%

%
\section{Introduction}
Medical diagnostics often involve the identification and analysis of various diagnostic indicators called biomarkers that can help understand the underlying pathology to inform patient care. When clinicians interpret point-of-care lung ultrasound (LUS), they identify biomarkers such as A-lines, B-lines, and pleural line thickness and then use these to inform diagnoses and patient care decisions. Training AI networks to specifically identify and classify these biomarkers can both enhance machine learning and render AI outputs to be relatable to clinicians.

Current typical ANN approaches train models end-to-end on the downstream task. By nature of this task specific training, the learnt features are task specific and do not readily generalize to other tasks. \cite{Zech2018VariableStudy} These task specific models require retraining or transfer learning to adapt to new end tasks. \cite{Tan2018OptimizeFine-Tuning}


We break down the diagnostic learning task into first predicting generic visual biomarkers using weak supervision followed by fitting task specific Expert models on these predicted biomarkers. Decoupling the feature learning from the downstream task enables the model to learn generic features that can readily adapt to new downstream tasks. Our approach of separating the biomarker detection and downstream analysis is analogous to the separation of the human visual cortex and prefrontal cortex of the brain.

The Biomarker detection model is trained to extract general-purpose LUS biomarker feature attributes, representing an entire video with a short vector of highly meaningful features. Compared to typical multi-layer-perceptron (MLP) classifiers that use 2-3 fully-connected (FC) layers, each with 1024-4096 features, our biomarker feature vector is much shorter, currently only 38 scalar values. Each feature carries specific meaning (e.g. the number of B-lines present in the video, which is a good indicator of the severity of the patient).  Having such a succinct representation allows easier training of subsequent AI models to perform a variety of end tasks, including future tasks which were not originally anticipated when training the biomarker model. These 38 biomarker features that capture the consolidated view of a video lack any spatial or temporal grounding of these features thus only providing a weak supervision to train the Biomarker model. This problem setup provides some potential benefits highlighted below.

\textbf{Reduced training cost:} The Biomarker model needs to be trained only once and can be easily adapted to any task by training a subsequent expert model on top. The expert model can now be simpler than the biomarker model as it operates on concise biomarker features rather than on image or video space (an extreme dimensionality reduction). Because the expert model is analyzing a much simpler, more apropos representation, it can be trained using less data (with the task-specific labels), whereas more data is needed to train the more complex Biomarker model. Thus the training cost of each expert model is lower.  Accordingly, small amounts of patient-specific or task-specific data might be used to create a new AI model by leveraging the biomarker model's capabilities (which might be learned from a private dataset that was only used to train the biomarker model).


\textbf{Annotation ease:} Human annotators can provide simple, discrete (binary) cues of (presumably) clinically significant features, making no use of tedious segmentation labels, thus reducing the annotation burden.   These visual semantic biomarker features are intended to be quickly assignable using binary checkboxes (often as a gestalt observation) by clinical experts.  We then leverage weak supervision to ultimately achieve 38 scalar biomarker features.

\textbf{Interpretable:} The predicted biomarkers are relatable to clinicians who can readily understand and verify the Biomarker model's outputs.  This may provide confidence and insight into the immediate inputs of the subsequent diagnostic classification, helping establish trust in the model. We hope our method provides a step towards overall model interpretability. \cite{Lipton2016TheInterpretability}


\textbf{Causal in nature:} By introducing this information bottleneck of training the expert model only on biomarker features, the expert model is limited from overfitting to spurious correlations in the data. \cite{Gordaliza2022TranslationalRepresentations} took a similar approach to train VAE's only using a restricted set of input latent variables thus preventing the model from overfitting to confounding variables.

In the rest of the paper, we introduce the biomarker features and show the ease with which our biomarker model can be adapted to diagnostic tasks.








\subsubsection{Related Work}

Current LUS AI approaches \cite{Roy2020DeepUltrasound,Born2021AcceleratingAnalysis,GareTheAI} train models directly for the diagnostic end task, sometimes using segmentation labels for pretraining \cite{Xue2021ModalityInformation,Gare2021DENSEDETECTION} to help improve performance. Although segmentation labels may be thought of as a type of biomarker, segmentations do not succinctly describe the key properties of the videos and segmentation labels are costly to obtain. 

The closest prior work may be \cite{Durrani2022AutomaticPneumonia}, which annotated video frames to indicate the presence (or absence) of few LUS biomarkers (consolidation, effusion) and trained networks to detect the same. Comparatively, our biomarker features capture a broader range of \textit{qualitative} characteristics, and our annotated biomarkers are gestalt'ed \textit{video} features (at low labeling cost) rather than per-frame features, thus providing a weak supervision for model training. We further use these rich but compact biomarker features for diagnostic tasks. 

Transfer learning is a popular technique used in the medical domain when working with limited training data, wherein a model trained on cross-domain or cross-modal data is fine-tuned for the intended task. \cite{Cheng2021TransferSegmentation,Tan2018OptimizeFine-Tuning} Biomarkers provide a potentially better pretraining task, learning generic features that can be adapted for related downstream tasks.



\textbf{Contributions} Our main technical contribution is (1) the proposed ``biomarker'' pre-task that helps generalize to pulmonary LUS diagnostic tasks. Other key contributions are: (2) We quantitatively define and classify LUS biomarker features, (3) We believe this to be the first work using ANN models on LUS to predict S/F ratio, and for (4) predicting diverse pulmonary diseases.

\section{Methodology}

\subsubsection{LUS Biomarkers}
Merriam-Webster defines a biomarker as ``a distinctive biological or biologically derived indicator of a process, event, or condition.'' Therefore an ideal biomarker is one that is both sensitive and specific for a disease process. Precise characterization of the set of biomarkers is necessary to optimize the biomarker's predictive power. For example, A-line count and intensity directly relate to normal lung while B-line count and intensity relate to disease lung. Likewise, pleural indents and breaks are biomarkers of disease. Thus, quantitative labeling of these biomarkers yields better test performance characteristics. To date, a few specific lung ultrasound biomarkers have been described, e.g. the “lung point” for pneumothorax and a sonolucent space between the parietal and visceral pleural surfaces for pleural effusion. \cite{Lichtenstein2019CurrentExperts}

In an effort to capture the qualitative characteristics of LUS diagnostic indicators, we quantitatively define and classify the biomarker features. We categorize the shape, size, location, and quantity of: A-line, B-line, Pleural line thickness \& location, Pleural indents \& breaks, Consolidation, and Effusion, \cite{Lichtenstein2019CurrentExperts} resulting in 38 (binary) biomarker features (refer to Figure \ref{fig:annotator} below and to Table~5 in supplementary material for the detailed list of biomarkers). The full list of biomarkers was developed by the authors achieving consensus among the senior clinicians. We recognize that the defined biomarkers aren't comprehensive and may need to be augmented to better generalize to future diagnostic tasks. 
%




In our experience, the use of an intuitive online quantitative annotator allowed trained sonographers to label and score hundreds of video clips in a fraction of the time that it would take to semantically label these same clips by drawing boundary masks.

\begin{figure}
\centering
\begin{minipage}[b]{\textwidth}
\includegraphics[width=\textwidth]{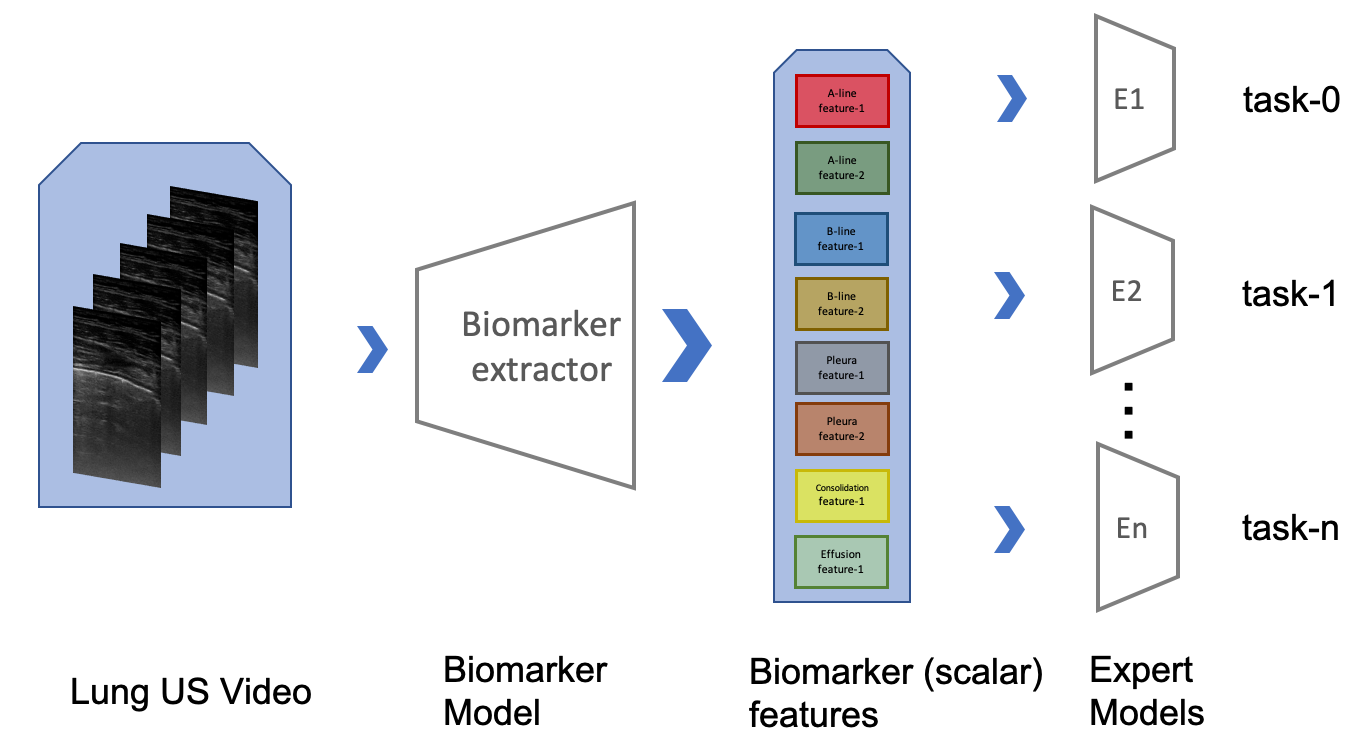}
\caption{
The Biomarker model condenses the video into  multi-purpose biomarker feature attributes on which light weight task specific \emph{expert} models are learnt with no loss in performance. 
} \label{fig:sytem_overview}
\end{minipage}
%
\begin{minipage}[b]{\textwidth}
\includegraphics[width=\textwidth]{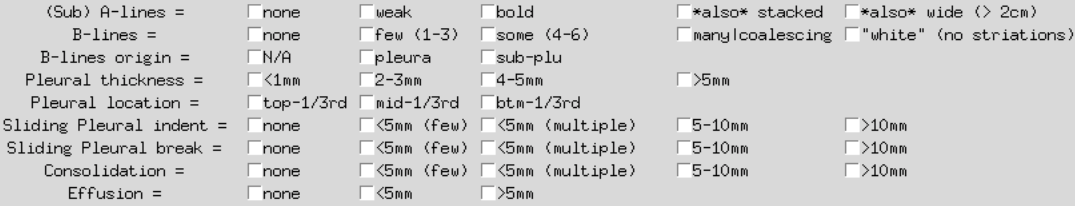}
\caption{The checkbox annotator used for labeling the 38 biomarkers.} 
\label{fig:annotator}
\end{minipage}
\end{figure}

\vspace{-1.0em}

\subsection{Data}

We compiled a lung ultrasound dataset of linear probe videos consisting of 189 patients (718 videos) with multiple (minimum 2 per patient) ultrasound B-scans of left and right lung regions at depths ranging from 4cm to 6cm under different scan settings, obtained using a Sonosite X-Porte ultrasound machine. (Fig. 3 in supplementary material highlights the various dataset distribution.)

\textbf{Biomarkers:} We labeled the 38 binary biomarker features of every video. All the manual labeling was performed by individuals with at least six months of training from a pulmonary ultrasound specialist.


\textbf{Lung-severity:} Each video was scored into 4 lung-severity classes by using the scoring scheme as defined in \cite{SimpleClinicalTrials.gov} and similarly used in \cite{Roy2020DeepUltrasound,GareTheAI}. The score-0 indicates a normal lung and scores 1 to 3 signify an abnormal lung with increasing level of severity. (Refer to Fig. 4 in supplementary material for sample images corresponding to the severity scores.)

\textbf{S/F ratio:} The S/F represents the ratio between measured blood oxyhemoglobin saturation (S) and the fraction of inspired oxygen (F). The lower this ratio, the more deranged is lung function. S/F is a standardized measurement used to assess lung function in research and at the bedside. We categorize the S/F ratio into the following 4-ranges: $[ > 430, 275-430, 180-275, <180 ]$. These reflect the maximum that the following devices are able to support lungs in oxygenating blood: ambient air, standard nasal cannula, venturi mask, and more intensive oxygen delivery devices only available in intensive care settings (e.g. non-invasive mechanical ventilation, invasive mechanical ventilation) respectively. 

\textbf{Disease:} The diagnosis for each patient was recorded and categorised into the following 7-categories: Healthy, COVID Pneumonia, Interstitial Lung Disease, Asthma/COPD Exacerbation, Cardiogenic Pulmonary Edema, Other lung diseases, Other non-lung diseases. These diseases are representative of common pulmonary pathologies encountered in the inpatient clinical setting of an academic American hospital. Each disease imposes disease-specific changes to the physical architecture of lung parenchyma, a property that clinicians use to diagnose pulmonary disease with ultrasound at the bedside.

\subsection{Model}

We carry out our experiments on the TSM network \cite{Lin2018TSM:Understanding} with ResNet-18 (RN18) \cite{He2016DeepRecognition} backbone and ImageNet \cite{DengImageNet:Database} pretrained weights. Similar to the setup in \cite{GareTheAI}, we use bi-directional residual shift with $1/8$ channels shifted in both directions.
%
%
%
We perform multi-clip evaluation similar to \cite{GareTheAI} wherein we evaluate n clips (n=4) and the mean prediction across the clips, and use the same frame sampling strategy wherein we feed the model 15 frame inputs selected from 15 equally spaced segments beginning with a random start frame. 

The model is trained on images which are resized to 224x224 pixels using bilinear interpolation. We augment the training data using standard augmentations:  horizontal flipping, pixel intensity scaling (scales $[0.8, 1.1]$), image-scaling ($\pm 20\%$), rotation ($\pm$ 15 degrees), and translation ($\pm$ 5\% pixels). The training set is oversampled to address the class imbalance \cite{Rahman2013AddressingDatasets}.
%
%

\subsubsection{Classifiers}
We use Decision Tree (DT), SVM, Random Forest (RF), AdaBoost (AB), Nearest Neighbours (NN), and MLP classifiers from the \href{https://scikit-learn.org/stable/supervised\_learning.html#supervised-learning}{scikit-learn} python library with default parameters. MLP Large uses 3 hidden layers (128, 64, 32) with adaptive learning rate. To handle the randomness in these simple classifiers we fit 3 models and choose the best model. 
Similar to the end-to-end trained model, we oversample the training data and perform multi-clip evaluation on the simple classifiers.



\subsection{Training Strategy}

\subsubsection{Implementation}
The network is implemented with PyTorch Lightning and trained using the stochastic gradient descent algorithm \cite{Bottou2010Large-scaleDescent} with an Adam optimizer \cite{Kingma2015Adam:Optimization} set with an initial learning rate of $0.0001$, to optimize over cross-entropy loss. The model is trained on an Nvidia RTX A6000 GPU, with a batch size of 4 for 100 epochs. The ReduceLRonPlateau learning rate scheduler is used, which reduces the learning rate by a factor (0.5) when the performance metric (accuracy) plateaus on the validation set. For the final evaluation, we pick the best model with highest validation set accuracy to test on the held-out test set.


\textbf{Metrics}
We report accuracy, precision, F1 score \cite{GareTheAI} and AUC of ROC metrics \cite{Kim2020ChangesStudy}, with weighted average (weights correspond to support of each class) and consider one-vs-one approach for the multi-class classification. \cite{Fawcett2006AnAnalysis}

\section{Experiments and Results}

\subsection{Cross-validation}
We randomly split the 189 patients into four equal folds, with one fold as the held-out test set (46 patients with 162 videos) and use the other three folds for training by performing \emph{3-fold cross-validation by training on two folds and validating on the other}. \emph{By creating patient based splits this provides a stricter evaluation} (recommended in \cite{Roshankhah2021InvestigatingImagesa}), although the folds don't share the same distribution of classes. We report the metrics on the held-out test set in the form of mean and standard deviation over the three independent cross-validation runs.

\subsection{Biomarker feature training}
We train the biomarker model to predict the 38 biomarker features using binary-cross-entropy loss and use the trained model for the diagnostic tasks by fitting classifier models on top. Refer to Table~\ref{tab:biomarker} for biomarker model metrics.


\subsection{Diagnostic classification task}
We train three task specific models end-to-end (E2E) for lung-severity (4 class), S/F ratio (4 class), and disease category (7 class) classification tasks respectively. We compare the performance of each with the biomarker model (trained once on biomarker feature prediction task) fitted with multiple simple classifiers.

Table~\ref{tab:classification-scores} shows the mean and standard deviation of the classification metrics obtained from 3-fold cross-validation runs on the independent held-out test set. We observe that the biomarker model achieves comparable performance (often surpassing) the end-to-end trained model on all three diagnostic tasks. By having been trained only once, the Biomarker model ready adapts to different downstream tasks at fractional training cost compared to task-specific end-to-end models, with no loss in performance.

The question that arises now is whether the performance gains of the biomarker model derive from the biomarker features themselves or instead from the different classifier heads (vs. the traditional FC MLP in an ANN). So, we evaluate by fitting simple classifiers on resnet features (from the layer before the last linear classification layer (having 512 embedding size)) of the end-to-end trained model similar to \cite{Almabdy2019DeepRecognition,LinBilinearRecognition}. Although this helps improve the end-to-end model's performance, it remains subpar to biomarker model's performance, thus indicative of the effectiveness of biomarker features.

We further evaluated if training an end-to-end model using biomarker pretrained weights improves performance for the lung-severity classification task and observed that it boosts the end-to-end model's performance and has comparable performance to the biomarker model (refer to section `Pretrain on Biomarkers' of Table~\ref{tab:classification-scores}).  We conclude that pretraining on biomarker features provides benefits in both scenarios, whether traditional pretraining or our explicit use by simple Expert models.


In another experiment, we evaluated the transfer-learning/adaptability of the biomarker model by comparing it with the lung-severity end-to-end trained model fit with simple classifiers to predict S/F ratio and Disease categories. We observe (refer to section `Lung-severity Features' of Table~\ref{tab:classification-scores}) that adapting the lung-severity model performs better than the end-to-end trained S/F ratio and disease task models. Yet, this approach sill falls short compared to the biomarker model, suggesting that Biomarker features are more adaptable.

In summary, the Biomarker model generalized better to different diagnostic tasks, and it matched or exceeded the performance of end-to-end trained models. 





\begin{table*}[h]
    \centering
    \caption{Classification metrics. The biomarker model trained only once for biomarker feature extraction and adapted for diagnostic tasks achieves better performance than directly training end-to-end.
    %
    }
    \label{tab:classification-scores}
    
    \resizebox{1.0\textwidth}{!}{

    \begin{tabular}{lcc|lcc|lccr}
        \toprule
        Method & \multicolumn{2}{c}{Lung-severity} & & \multicolumn{2}{c}{S/F ratio}  & & \multicolumn{2}{c}{Disease} & Train Time \\ 
        
        \cmidrule{2-3} \cmidrule{5-6} \cmidrule{8-9} \\[-\normalbaselineskip]
        
         & AUC of ROC & Accuracy  && AUC of ROC & Accuracy && AUC of ROC & Accuracy &  (in min) \\
        \cmidrule{2-3} \cmidrule{5-6} \cmidrule{8-9} \\[-\normalbaselineskip]
        
        \midrule
        \midrule
        
        \multicolumn{10}{c}{End-to-end} \\
        
        \midrule

        E2E & 
        81.5 $\pm$ 1.0 & 57.6 $\pm$ 2.8 &&
        54.6 $\pm$ 4.3 & 29.6 $\pm$ 1.5 &&
        59.2 $\pm$ 1.8 & 29.2 $\pm$ 2.3 &
        3 x 194 \\ 
        
        \cmidrule{2-3} \cmidrule{5-6} \cmidrule{8-9} \\[-\normalbaselineskip]
        
        DT & 
        68.2 $\pm$ 2.0 & 55.6 $\pm$ 2.8 &&
        52.9 $\pm$ 2.8 & 32.1 $\pm$ 5.0 &&
        51.8 $\pm$ 2.0 & 26.5 $\pm$ 1.0 &
        3 x 1 \\ 
        
        SVM & 
        80.5 $\pm$ 1.9 & 57.8 $\pm$ 2.3 &&
        55.4 $\pm$ 2.5 & 31.3 $\pm$ 4.7 &&
        58.2 $\pm$ 0.4 & 26.9 $\pm$ 0.8 &
        3 x 1 \\
        
        RF & 
        81.3 $\pm$ 0.2 & 59.1 $\pm$ 0.8 &&
        56.1 $\pm$ 1.7 & 34.8 $\pm$ 2.5 &&
        58.7 $\pm$ 1.0 & 29.0 $\pm$ 1.3 &
        3 x 2 \\
        
        AB & 
        73.2 $\pm$ 4.8 & 50.4 $\pm$ 8.6 &&
        54.4 $\pm$ 3.2 & 31.3 $\pm$ 2.9 &&
        51.9 $\pm$ 2.8 & 18.5 $\pm$ 10.3 &
        3 x 2 \\
        
        NN & 
        74.6 $\pm$ 0.7 & 56.0 $\pm$ 3.0 &&
        52.5 $\pm$ 0.9 & 26.8 $\pm$ 3.6 &&
        54.6 $\pm$ 1.7 & 28.8 $\pm$ 2.8 &
        3 x 1 \\
        
        MLP & 
        80.3 $\pm$ 0.8 & 59.9 $\pm$ 1.3 &&
        54.2 $\pm$ 2.3 & 32.7 $\pm$ 3.1 &&
        57.7 $\pm$ 1.8 & \textbf{29.2} $\pm$ \textbf{1.5} &
        3 x 2 \\
        
        MLP large & 
        79.5 $\pm$ 1.7 & 59.3 $\pm$ 1.8 &&
        54.0 $\pm$ 3.1 & 34.6 $\pm$ 2.0 &&
        57.2 $\pm$ 0.8 & 28.6 $\pm$ 1.8 &
        3 x 3 \\
 
        \midrule
        \midrule
        
        \multicolumn{9}{c}{Biomarker Features} &
        \textbf{1 x} 173.21 \\ 
        
        \midrule
        
        
        
        \cmidrule{2-3} \cmidrule{5-6} \cmidrule{8-9} \\[-\normalbaselineskip]
        
        DT & 
        68.3 $\pm$ 2.5 & 54.1 $\pm$ 4.5 &&
        55.3 $\pm$ 1.7 & 35.2 $\pm$ 2.7 &&
        53.7 $\pm$ 2.3 & 24.7 $\pm$ 2.2 &
        3 x 1 \\ 
        
        SVM & 
        82.9 $\pm$ 1.3 & 61.3 $\pm$ 2.1 &&
        \textbf{61.7} $\pm$ \textbf{2.2} & 35.8 $\pm$ 4.1 &&
        \textbf{64.2} $\pm$ \textbf{3.7} & 24.1 $\pm$ 1.5 &
        3 x 1 \\
        
        RF & 
        81.2 $\pm$ 1.7 & 60.3 $\pm$ 3.2 &&
        60.5 $\pm$ 2.1 & 38.5 $\pm$ 1.9 &&
        61.2 $\pm$ 3.0 & 28.4 $\pm$ 0.9 &
        3 x 2 \\
        
        AB & 
        72.7 $\pm$ 2.3 & 50.8 $\pm$ 2.0 &&
        59.3 $\pm$ 3.7 & \textbf{40.3} $\pm$ \textbf{4.8} &&
        56.2 $\pm$ 3.6 & 17.9 $\pm$ 5.3 &
        3 x 2 \\
        
        NN & 
        76.2 $\pm$ 3.4 & 54.1 $\pm$ 2.5 &&
        52.8 $\pm$ 2.5 & 29.0 $\pm$ 3.5 &&
        54.5 $\pm$ 5.2 & 20.8 $\pm$ 3.4 &
        3 x 1 \\
        
        MLP & 
        \textbf{83.2} $\pm$ \textbf{1.4} & 62.1 $\pm$ 2.0 &&
        61.0 $\pm$ 1.7 & 37.4 $\pm$ 1.3 &&
        63.7 $\pm$ 3.5 & 25.3 $\pm$ 2.0 &
        3 x 2 \\
        
        MLP large & 
        81.6 $\pm$ 0.9 & 61.1 $\pm$ 1.5 &&
        57.4 $\pm$ 1.7 & 36.6 $\pm$ 4.1 &&
        61.1 $\pm$ 4.4 & 29.2 $\pm$ 1.8 &
        3 x 3 \\

        \midrule
        \midrule
        
        & \multicolumn{3}{c}{Pretrain on Biomarkers} 
        & \multicolumn{5}{c}{Lung-severity Features} & \\
         
        \midrule
        
        \cmidrule{2-3} \cmidrule{5-6} \cmidrule{8-9} \\[-\normalbaselineskip]
        
        E2E & 
        82.4 $\pm$ 2.0 & 57.4 $\pm$ 0.5 &&
         &  &&
         &  &
        - \\ 
        
        \cmidrule{2-3} \cmidrule{5-6} \cmidrule{8-9} \\[-\normalbaselineskip]
        
        DT & 
        68.7 $\pm$ 3.4 & 57.8 $\pm$ 3.7 &&
        56.4 $\pm$ 1.7 & 37.2 $\pm$ 1.6 &&
        53.7 $\pm$ 0.6 & 22.6 $\pm$ 2.0 &
        3 x 1 \\ 
        
        SVM & 
        81.9 $\pm$ 2.7 & 60.5 $\pm$ 3.6 &&
        60.2 $\pm$ 1.9 & 33.5 $\pm$ 2.6 &&
        59.3 $\pm$ 2.7 & 20.2 $\pm$ 2.8 &
        3 x 1 \\
        
        RF & 
        81.6 $\pm$ 3.3 & 61.1 $\pm$ 2.5 &&
        58.2 $\pm$ 1.2 & 38.9 $\pm$ 2.2 &&
        59.4 $\pm$ 1.8 & 26.1 $\pm$ 2.0 &
        3 x 2 \\
        
        AB & 
        71.2 $\pm$ 3.4 & 47.9 $\pm$ 13.8 &&
        56.8 $\pm$ 3.4 & 38.3 $\pm$ 3.5 &&
        56.6 $\pm$ 1.8 & 16.9 $\pm$ 4.1 &
        3 x 2 \\
        
        NN & 
        76.5 $\pm$ 2.4 & 61.1 $\pm$ 0.5 &&
        56.4 $\pm$ 0.8 & 28.8 $\pm$ 3.0 &&
        53.7 $\pm$ 3.5 & 17.5 $\pm$ 2.8 &
        3 x 1 \\
        
        MLP & 
        82.6 $\pm$ 2.3 & 61.7 $\pm$ 3.6 &&
        59.7 $\pm$ 2.6 & 37.0 $\pm$ 2.0 &&
        60.1 $\pm$ 1.7 & 25.7 $\pm$ 2.0 &
        3 x 2 \\
        
        MLP large & 
        80.6 $\pm$ 2.1 & \textbf{62.6} $\pm$ \textbf{2.0} &&
        57.9 $\pm$ 2.9 & 37.2 $\pm$ 2.6 &&
        59.5 $\pm$ 1.6 & 25.9 $\pm$ 1.3 &
        3 x 3 \\

        
        \bottomrule
        
    \end{tabular}
    
    } 
    
\end{table*}

\textbf{POCOVID Dataset:}
We next tested the lung-severity models on the 21 linear probe videos in the publicly usable subset of the POCOVID-Net dataset. \cite{Born2021AcceleratingAnalysis}  We observed (refer to Table~\ref{tab:pocovid}) that the biomarker (Bio) model generalizes better than the end-to-end (E2E) trained model.


\begin{table}[!htb]
    \begin{minipage}{.5\linewidth}
    \centering
     \caption{Biomarker feature model metrics.}
    \label{tab:biomarker}
    \resizebox{0.8\textwidth}{!}{
    \begin{tabular}{lccc}
        \toprule
        Method & Accuracy & Precision & F1-score\\
        \midrule
  Biomarker & 76.4 $\pm$ 1.0 & 70.8 $\pm$ 0.1 & 73.0 $\pm$ 0.5 \\
        \bottomrule
    \end{tabular}
    }
    \vspace{15px}
\centering
\caption{POCOVID Dataset results.}
\label{tab:pocovid}
 \begin{tabular}{lcc}
        \toprule
        Method  & AUC of ROC & Accuracy  \\

        \midrule
        
        E2E & 61.5 $\pm$ 3.2 & 39.7 $\pm$ 4.5 \\ 


 E2E (SVM) &  58.8 $\pm$ 2.0 & 36.5 $\pm$ 2.2  \\ 
 E2E (MLP) &  58.9 $\pm$ 0.7 & 39.7 $\pm$ 2.2 \\ 
 
        
        
        
 Bio (SVM) &  65.6 $\pm$ 3.1 & 25.4 $\pm$ 5.9  \\ 
 Bio (MLP) &  65.9 $\pm$ 4.1 & 30.2 $\pm$ 2.2 \\ 
 
        \bottomrule
    \end{tabular}
    \end{minipage}%
    \begin{minipage}{.5\linewidth}
      \centering
        
 \caption{Inter-labeler agreement with LUS Specialist.
    }
    \label{tab:inter_labeler}
    
    \begin{tabular}{lcc}
        \toprule
        Labeler/Model & AUC of ROC & Accuracy \\

        \midrule
 
 trainee  & - & 56.79 \\ 
 E2E  & 75.34 & 43.21 \\ 
 E2E (SVM)  & 70.05 & 46.30 \\ 
 E2E (MLP)  & 73.70 & 45.06 \\ 
 Bio (SVM)  & 76.38 & 47.53 \\ 
 Bio (MLP)  & 77.64 & 49.38  \\ 

        \bottomrule
    \end{tabular}
    \end{minipage} 
\end{table}

\textbf{Inter-labeler agreement:}
We compare the lung-severity model's predictions and the trainee's labels (used for training) with labels obtained from an LUS specialist on the test set. We observe (refer to Table~\ref{tab:inter_labeler}) the trainee has 56\% agreement with the specialist and the biomarker (Bio) model has higher agreement than the end-to-end (E2E) model.

\section{Conclusion}

Biomarkers can provide improved pre-training and supervision tasks, which helped train a robust lung-ultrasound AI model that can generalize to other tasks.
In the future, we would like to incorporate self-supervised training into our biomarker model and to explore ways the model can self-discover interpretable biomarker features rather than pre-specifying them.
%
%
%
\bibliographystyle{splncs04}
\bibliography{references}


\begin{figure}
\includegraphics[width=\textwidth]{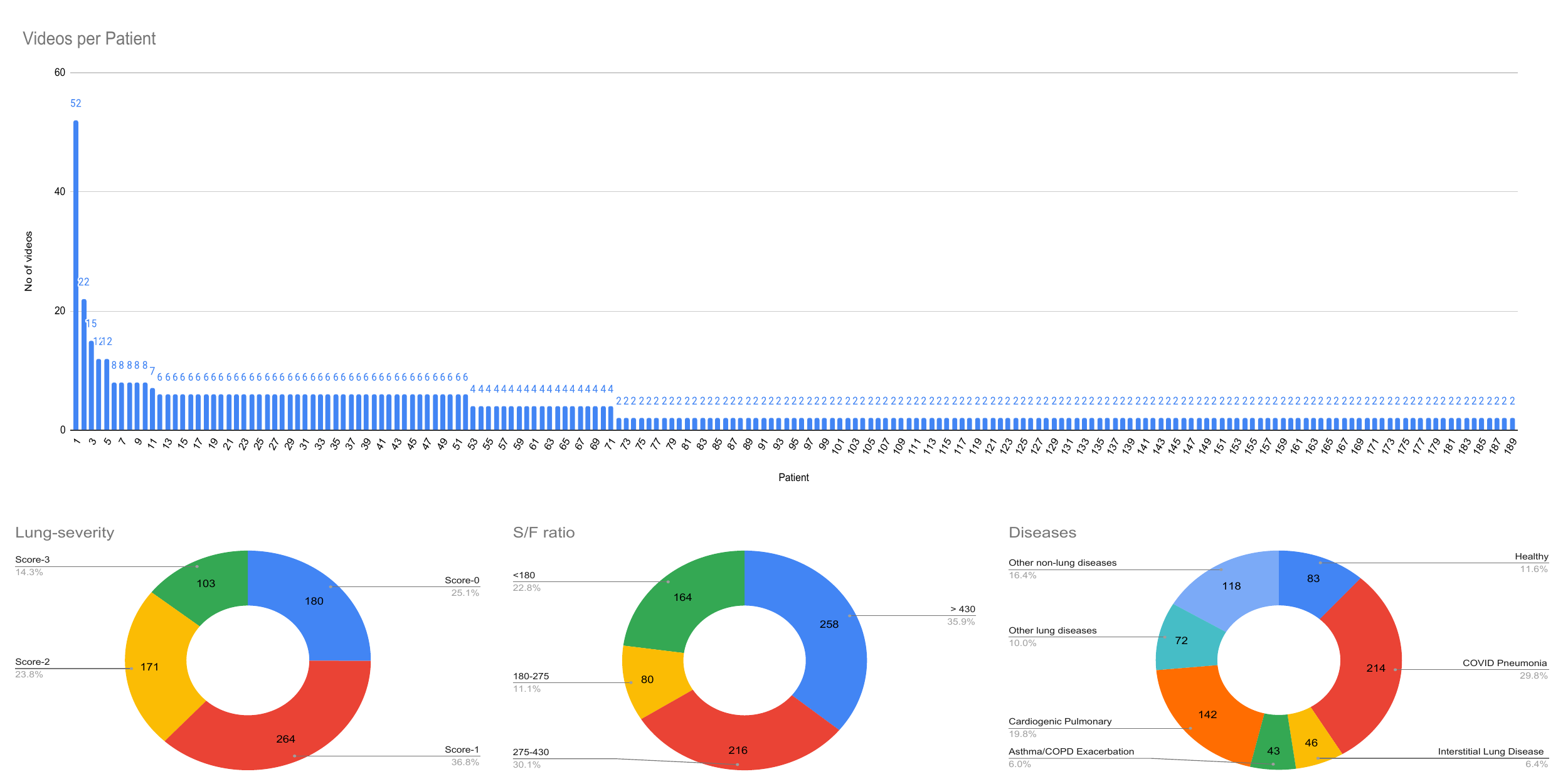}
\caption{
The various dataset distributions. (top) Videos per patient; (bottom) Videos per Lung-severity, S/F ratio, and Disease categories from left to right respectively.
} \label{fig:sytem_overview}
\end{figure}



\begin{table}[]
\centering
    \caption{Biomarkers features (and sub-categories) used in the paper that try to capture the qualitative characteristics of the biomarkers by grouping them into discretized ranges/classes.
    }
\label{tab:diagnostic_disease_property}

\resizebox{1.0\textwidth}{!}{

\begin{tabular}{|l|l|c|}
\hline
Biomarker Feature & Captured qualitative characteristics & Categories\\
\hline
        
A-line  & signifies A-line strength and spread  & 5 \\ \hline
B-line  & signifies number of B-lines and type (coalescing or white lung \cite{Lichtenstein2019CurrentExperts}) & 5 \\ \hline
B-line origin   & signifies whether B-lines originate at pleura or sub-pleura  & 3 \\ \hline
Pleural line thickness & signifies the thickness of the pleural line & 4  \\ \hline
Pleural line location   & signifies the pleural line location in the image & 3 \\ \hline
Pleural indents & signifies the indentations in the pleural line  & 5 \\ \hline
Pleural breaks  & signifies the breaks in the pleural line & 5 \\ \hline
Consolidation   & signifies the size of consolidation & 5 \\ \hline
Effusion    & signifies the size of effusion & 3 \\ \hline
\end{tabular}

} 

\end{table}


\newlength{\width}
\setlength{\width}{0.35 in}
\newlength{\height}
\setlength{\height}{0.25 in}

\begin{figure}[!ht] 
\centering

\setlength{\tabcolsep}{1pt} 
\def\arraystretch{0.3} 


\resizebox{0.9\columnwidth}{!}{
\begin{tabular}{cccc}

\tiny Score-0 &
\tiny Score-1 &
\tiny Score-2 &
\tiny Score-3 \\

\includegraphics[height = \height, width = \width]{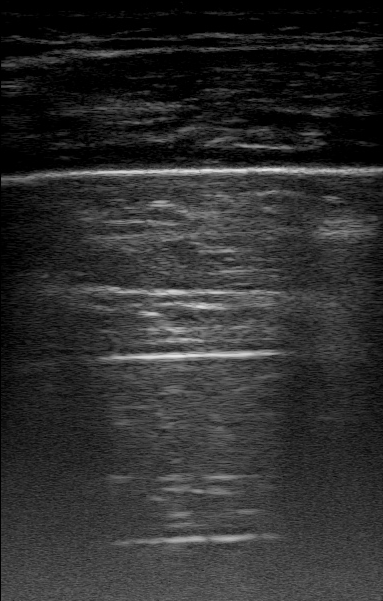} &
\includegraphics[height = \height, width = \width]{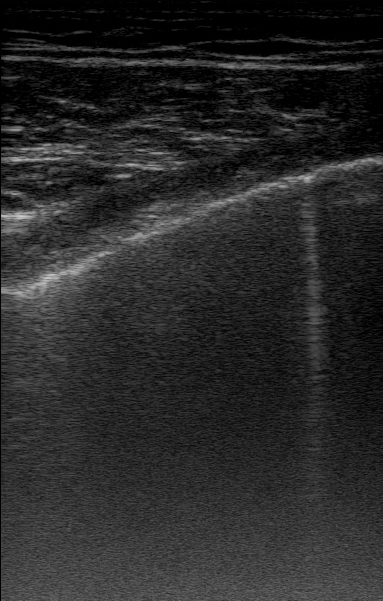} &
\includegraphics[height = \height, width = \width]{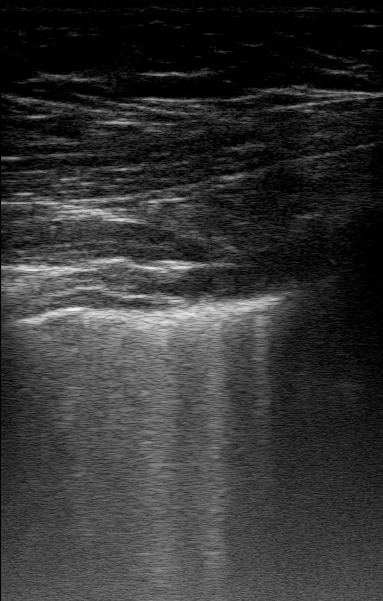} &
\includegraphics[height = \height, width = \width]{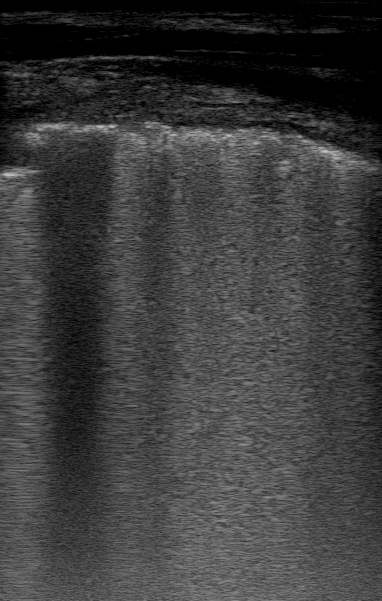} \\

\end{tabular}

}

\caption{
\small Sample test set video images corresponding to the four lung-severity classes. 
}
\label{fig:datasest_images}
\end{figure}


\begin{table*}[h]
    \centering
    \caption{Classification metrics using I3D Model. We make use of single stream I3D model \cite{Carreira2017QuoDatasetb} with ResNet-50 backbone with Kinetics-400 pretrain weights from \href{https://pytorchvideo.readthedocs.io/en/latest/model\_zoo.html}{PyTorch Moodel Zoo} to train the three end-to-end (E2E) models and the biomarker model. We again observe that with the new I3D model architecture, the biomarker model trained only once for biomarker feature extraction and adapted for diagnostic tasks achieves better performance than directly training end-to-end.
    Highest scores are shown in bold.
    }
    \label{tab:i3d_model}
    
    \resizebox{1.0\textwidth}{!}{

    \begin{tabular}{lcc|lcc|lccr}
        \toprule
        Method & \multicolumn{2}{c}{Lung-severity} & & \multicolumn{2}{c}{S/F ratio}  & & \multicolumn{2}{c}{Disease} & Train Time \\ 
        
        \cmidrule{2-3} \cmidrule{5-6} \cmidrule{8-9} \\[-\normalbaselineskip]
        
         & AUC of ROC & Accuracy  && AUC of ROC & Accuracy && AUC of ROC & Accuracy &  (in min) \\
        \cmidrule{2-3} \cmidrule{5-6} \cmidrule{8-9} \\[-\normalbaselineskip]
        
        \midrule
        \midrule
        
        \multicolumn{10}{c}{End-to-end} \\
        
        \midrule

        E2E & 
        81.1 $\pm$ 2.9 & 57.2 $\pm$ 3.8 &&
        56.2 $\pm$ 1.3 & 34.9 $\pm$ 4.7 &&
        59.1 $\pm$ 1.2 & 24.9 $\pm$ 3.8 &
        3 x 263 \\

        \midrule
        \midrule
        
        \multicolumn{9}{c}{Biomarker Features} &
        \textbf{1 x} 256 \\ 
        
        \midrule
        
        \cmidrule{2-3} \cmidrule{5-6} \cmidrule{8-9} \\[-\normalbaselineskip]
        
        DT & 
        65.2 $\pm$ 3.6 & 51.6 $\pm$ 3.4 &&
        54.6 $\pm$ 1.5 & 33.3 $\pm$ 3.8 &&
        52.3 $\pm$ 1.5 & 22.4 $\pm$ 1.3 &
        3 x 1 \\ 
        
        SVM & 
        82.4 $\pm$ 0.4 & 58.2 $\pm$ 1.6 &&
        61.6 $\pm$ 1.5 & 36.8 $\pm$ 1.2 &&
        64.9 $\pm$ 4.5 & 26.1 $\pm$ 1.8 &
        3 x 1 \\
        
        RF & 
        81.7 $\pm$ 0.1 & 60.5 $\pm$ 2.3 &&
        \textbf{61.8} $\pm$ \textbf{1.5} & \textbf{40.5} $\pm$ \textbf{1.8} &&
        \textbf{65.3} $\pm$ \textbf{3.2} & \textbf{30.2} $\pm$ \textbf{4.0} &
        3 x 2 \\
        
        AB & 
        71.9 $\pm$ 5.6 & 48.4 $\pm$ 3.9 &&
        56.1 $\pm$ 1.9 & 34.4 $\pm$ 5.8 &&
        59.1 $\pm$ 3.7 & 20.8 $\pm$ 2.5 &
        3 x 2 \\
        
        NN & 
        72.8 $\pm$ 0.3 & 51.6 $\pm$ 0.6 &&
        54.7 $\pm$ 1.2 & 31.1 $\pm$ 0.8 &&
        58.4 $\pm$ 1.9 & 23.0 $\pm$ 2.1 &
        3 x 1 \\
        
        MLP & 
        \textbf{82.8} $\pm$ \textbf{1.1} & 60.7 $\pm$ 1.3 &&
        59.9 $\pm$ 0.4 & 37.4 $\pm$ 2.5 &&
        64.2 $\pm$ 3.6 & 25.7 $\pm$ 1.2 &
        3 x 2 \\
        
        MLP large & 
        80.3 $\pm$ 1.7 & \textbf{61.3} $\pm$ \textbf{1.8} &&
        57.6 $\pm$ 1.6 & 36.2 $\pm$ 2.3 &&
        62.2 $\pm$ 3.6 & 27.6 $\pm$ 2.6 &
        3 x 3 \\
    
        
        \bottomrule
        
    \end{tabular}
    
    } 
    
\end{table*}

\vspace{-10.0em}

\begin{table*}[h]
    \centering
    \caption{Detailed classification metrics for the end-to-end (E2E) model and the biomarker model on Lung-severity task (As additional metric for lung-severity task sub-part of Table~\ref{tab:classification-scores}).
    }
    \label{tab:lung-severity}
    
    \begin{tabular}{lc|c|c|c}
        \toprule
        Method & AUC of ROC & Accuracy & Precision & F1-score \\

        \midrule
        \midrule
        
        \multicolumn{5}{c}{End-to-end} \\
        
        \midrule
        
E2E & 81.53 $\pm$ 1.02 & 57.61 $\pm$ 2.78 & 58.42 $\pm$ 2.50 & 57.01 $\pm$ 3.16 \\ 
 
        \midrule

 DT  & 68.18 $\pm$ 2.02 & 55.56 $\pm$ 2.81 & 55.65 $\pm$ 2.82 & 55.27 $\pm$ 2.98 \\ 
 SVM  & 80.53 $\pm$ 1.90 & 57.82 $\pm$ 2.27 & 58.46 $\pm$ 2.24 & 57.47 $\pm$ 1.99 \\ 
 RF  & 81.28 $\pm$ 0.23 & 59.05 $\pm$ 0.77 & 59.52 $\pm$ 0.46 & 58.41 $\pm$ 0.60 \\ 
 AB  & 73.17 $\pm$ 4.78 & 50.41 $\pm$ 8.59 & 47.47 $\pm$ 10.66 & 45.07 $\pm$ 11.79 \\ 
 NN  & 74.58 $\pm$ 0.71 & 55.97 $\pm$ 3.04 & 57.06 $\pm$ 3.36 & 55.97 $\pm$ 3.16 \\ 
 MLP  & 80.29 $\pm$ 0.78 & 59.88 $\pm$ 1.33 & 60.19 $\pm$ 1.59 & 59.48 $\pm$ 0.94 \\ 
 MLP Large  & 79.49 $\pm$ 1.71 & 59.26 $\pm$ 1.82 & 59.46 $\pm$ 2.06 & 58.85 $\pm$ 1.37 \\ 
        
        \midrule
        \midrule
        
        \multicolumn{5}{c}{Biomarker Features} \\

        \midrule
        
 DT  & 68.26 $\pm$ 2.50 & 54.12 $\pm$ 4.52 & 54.55 $\pm$ 4.59 & 54.12 $\pm$ 4.52 \\ 
 SVM  & 82.93 $\pm$ 1.32 & 61.32 $\pm$ 2.10 & 62.12 $\pm$ 2.14 & 61.43 $\pm$ 2.06 \\ 
 RF  & 81.24 $\pm$ 1.73 & 60.29 $\pm$ 3.24 & 61.10 $\pm$ 3.15 & 60.27 $\pm$ 3.27 \\ 
 AB  & 72.72 $\pm$ 2.27 & 50.82 $\pm$ 2.04 & 55.46 $\pm$ 2.53 & 48.80 $\pm$ 1.26 \\ 
 NN & 76.25 $\pm$ 3.36 & 54.12 $\pm$ 2.49 & 54.58 $\pm$ 2.14 & 54.15 $\pm$ 2.34 \\ 
 MLP  & 83.24 $\pm$ 1.41 & 62.14 $\pm$ 2.04 & 62.83 $\pm$ 1.70 & 62.16 $\pm$ 2.10 \\ 
 MLP Large & 81.64 $\pm$ 0.93 & 61.11 $\pm$ 1.51 & 62.03 $\pm$ 1.27 & 61.20 $\pm$ 1.25 \\ 
        
        \bottomrule
    \end{tabular}
\end{table*}


\end{document}